\documentclass[11pt,leqno]{article}
\usepackage{amsfonts,latexsym}
\usepackage{amsthm}
\usepackage{amsmath}
\usepackage[dvips]{graphicx}
\usepackage{graphicx}
\usepackage{amscd}
\usepackage{color}
\usepackage{amssymb}
\usepackage{a4wide}

\newtheorem{theorem}{Theorem}

\numberwithin{equation}{section}
\newcommand{\Sfrac}[2]{{ \textstyle \frac{#1}{#2}}}

\newcommand{\tx}{\tilde{x}}
\newcommand{\tz}{\tilde{z}}
\newcommand{\ttt}{\tilde{t}}

\newcommand{\tzeta}{\tilde{\zeta}}
\newcommand{\txi}{\tilde{\xi}}
\newcommand{\tu}{\tilde{u}}

\newcommand{\tphi}{\tilde{\phi}}

\newcommand{\R}{{\mathbb{R}}}
\newcommand{\scrO}{{\mathcal{O}}}

\newcommand{\ve}{{\varepsilon}}

\newcommand{\dtt}{\frac{\partial}{\partial t}}
\newcommand{\dxx}{\frac{\partial}{\partial x}}

\newcommand{\dx}{\partial_x}
\newcommand{\dz}{\partial_z}
\newcommand{\dt}{\partial_t}

\newtheorem{thm}{Theorem}

\newcommand{\vbarWW}{\bar{V}}

\newtheorem{definition}{Definition}

\makeatletter
\renewcommand\theparagraph    {\@roman\c@paragraph}
\renewcommand \thesection {\@arabic\c@section}%{\@arabic\c@section}

\renewcommand\paragraph{\@startsection{paragraph}{4}{\z@}{1.5ex\@plus1ex\@minus.2ex}{0pt}{\normalfont\normalsize\slshape}}
\renewcommand\section{\@startsection {section}{1}{\z@}{-2.1ex \@plus -1ex \@minus -.2ex}{0.5ex\@plus.1ex}{\bfseries\Large}}

\title{A Mathematical Justification of the Momentum Density Function Associated to the KdV Equation}

\author{Samer Israwi\footnote{\texttt{s\_israwi83@hotmail.com}} \\
{\small Lebanese University, Faculty of Sciences 1} \\
{\small Hadath-Beirut, Lebanon} \\ \\
Henrik Kalisch\footnote{\texttt{henrik.kalisch@uib.no}} \\
{\small Department of Mathematics, University of Bergen} \\
{\small Postbox 7800, 5020 Bergen, Norway} 
}
\date{}

\begin{document}
\maketitle

\begin{abstract}
Consideration is given to the KdV equation as an approximate model for
long waves of small amplitude at the free surface of an inviscid fluid.
It is shown that there is an approximate momentum density associated to the KdV
equation, and the difference between this density and the physical momentum density
derived in the context of the full Euler equations
can be estimated in terms of the long-wave parameter.\\
\end{abstract}

\section{Introduction}
In the present contribution, we consider the question of momentum conservation
in the context of the Korteweg-de Vries equation.
The KdV equation 
\begin{align}\label{KdV}
\eta_{t}+ \eta_{x} + \ve \frac{3}{2} \eta \eta_{x}
                    + \mu \frac{1}{6} \eta_{xxx} = 0
\end{align}
is known to yield a valid description of surface waves 
for waves of small amplitude and large wavelength
at the free surface of an incompressible, inviscid fluid 
running in a narrow open channel where transverse effects can be neglected.

Suppose $h_0$ is the depth of the undisturbed fluid,
and let $\lambda$ denote a typical wavelength and by $a$ a typical amplitude
of a wavefield to be described. 
The nondimensional number $\ve = a/h_0$ then represents the relative amplitude.
If we define the long-wave parameter by $\mu = h_0^2 / \lambda^2$,
then the KdV equation is known to be a good model for waves at the free surface
of a fluid if the relations $\mu << 1$ and $\ve = \scrO(\mu)$.
The approximation can be made rigorous using the
techniques developed in \cite{BCL,Craig,Israwi,LannesBOOK,SchneiderWayne} and others.

It is well known that the KdV equation has an infinite number of formally
conserved integrals (indeed the conservation can be made rigorous by following
the work of \cite{BonaSmith}). 
If the equation is given in the non-dimensional form \eqref{KdV},
the first three conserved integrals are
\begin{align}
\label{conslaws}
\int_{- \infty}^{\infty} \eta \, dx,
\qquad
\int_{- \infty}^{\infty} \eta^2 \, dx,
\quad \mbox{ and } \quad
\int_{- \infty}^{\infty} \left( \Sfrac{\mu}{3 \ve} \eta_x^2  -  \eta^3 \right) \, dx.
\end{align}
The first integral is found to be invariant with respect to time $t$
as soon as it is recognized that the KdV equation can be written in
the form
\begin{align}\label{kdv_mass}
\dtt \left( \eta \right) + 
\dxx \Big( \eta + \ve \frac{3}{4} \eta^2 + \mu \frac{1}{6} \eta_{xx} \Big) = 0,
\end{align}
where the quantity appearing under the time derivative is interpreted 
as excess mass density, and the term appearing under the spatial
derivative is the mass flux through a cross section of unit width
due to the passage of a surface wave.
The second and third integral are sometimes called momentum and energy,
but this terminology may be misleading since these integrals
are not readily interpreted as approximations of the
physical momentum and energy appearing in the context of the Euler equations.
Indeed, the authors of \cite{AS} already state clearly that they do not
believe these integrals to be approximations of the physical momentum and density,
and further doubt was cast on this interpretation in more recent work \cite{AK1,KRI2015,KRIR2017}.

On the other hand, in physical flow problems, mass flux is often identical with momentum
density, so one might think that the term 
$\eta + \ve \frac{3}{4} \eta^2 + \mu \frac{1}{6} \eta_{xx}$
in \eqref{kdv_mass} might be interpreted as momentum flux.
This is indeed correct as shown in the recent work \cite{AK4}, 
where based on ideas developed in \cite{AK2},
it was shown how to find integral quantities that do represent
approximations to the physically relevant momentum and energy densities.
In particular, following the procedure laid out in \cite{AK4} gives the expression 
for momentum density as
\begin{equation}\label{MomDensityI}
I = \eta + \ve \frac{3}{4} \eta^2 + \mu \frac{1}{6}\eta_{xx}.
\end{equation}

Since the analysis in \cite{AK4} was based on a formal asymptotic analysis,
the question of whether this identity can be made mathematically rigorous
has so far remained open. In the present work we will prove that 
a firm mathematical proof can indeed be given.
The main result to be proved thus states that the density $I$ converges to
the physical momentum density defined in terms of a solutions of the governing
Euler equation for a perfect fluid if $\mu$ and $\ve$ tend to zero.
The precise statement is as follows.

\begin{thm} \label{Momdensity}
Let $(\zeta,\phi)$ be a solution of the water-wave problem defined below,
with initial data given by $(\zeta_0,\phi_0)\in (H^s(\R))^2$ for $s$ large enough.
Let $\eta$ be a solution of the KdV equation \eqref{KdV} 
with initial data $\eta_0=\zeta_0$.
Then there exists a constant depending on $s$, so that we have the estimate
\begin{equation}
\Big\| \int_{-1}^{\ve \zeta(\cdot,t)}\dx \phi(z,\cdot,t)dz 
- \eta(\cdot,t) - \ve \frac{3}{4} \eta^2(\cdot,t) - \mu \frac{1}{6}\eta_{xx}(\cdot,t) 
\Big\|_{L^{\infty}} \leq C \mu^2 (1+t).
\end{equation}
\end{thm}

% *****************************************************************************************
\section{Auxiliary results}
% *****************************************************************************************
% *****************************************************************************************
%
%
Denoting the original (dimensional) variables with a tilde,
we introduce a scaling to make the small amplitude
and long wavelength relative to the undisturbed depth
explicit. Thus we define new variables (without a tilde) by
%
%\begin{equation*}
%      \tx=\frac{x}{\lambda}, \quad \tz=\frac{z}{h_0},
%\quad \teta=\frac{\eta}{a}, \quad \ttt= \frac{c_0}{\lambda} t, 
%\quad \tphi= \frac{a \lambda g}{c_0} \phi 
%\end{equation*}
%\begin{equation*}
$      \tx = \lambda x,\ \tz = h_0 z, \ \tzeta = a \zeta, \ \ttt= \frac{\lambda}{c_0} t, 
\ \tphi= \frac{a \lambda g}{c_0} \phi.
$
%\end{equation*}
%%
Then we obtain the system
\begin{equation}\label{ww}
	\left\lbrace
	\begin{array}{lcl}
	\mu\dx^2\phi+\dz\phi^2=0 &\mbox{ in }&\Omega_t,\\
	\dz\phi=0,&\mbox{ at }& z=-1,\\
	\dt \zeta-\frac{1}{\mu}(-\mu\dx\zeta\dx\phi+\dz\phi)=0
	&\mbox{ at }& z= \ve \zeta,\\
	\dt\phi+\zeta+\frac{\ve}{2}(\dx\phi)^2+\frac{\ve}{2\mu}(\dz\phi)^2=0
        &\mbox{ at }& z= \ve\zeta,
	\end{array}\right.
\end{equation}
where $\Omega_t=\{(x,z),-1<z<\ve \zeta(x,t)\}$ is the fluid domain
delimited by the free surface $\{z=\zeta(x,t)\}$,
and the flat bottom $\{z=-1\}$, and where
$\phi(x,z,t)$, defined on $\Omega_t$ is the velocity potential
associated to the flow (that is, the two-dimensional 
velocity field ${\bf v}$ is given by ${\bf v}=(\dx\phi,\dz\phi)^T$).
As is well known, the existence of the velocity potential is guaranteed
by the assumption of irrotational flow.
The equations above are formally equivalent to the Zakharov-Craig-Sulem (ZCS) equations.
They are written in terms of the trace $\Phi(x,t) = \phi(x,\zeta(x,t),z)$ and
the Dirichlet-Neumann operator $G(\zeta)$ as
\begin{align}\label{Z-C-S}
\left\lbrace
\begin{array}{ll}
& \zeta_t - \frac{1}{\mu} G(\ve\zeta) \Phi = 0, \\
& \Phi_t +  \zeta + \frac{\ve}{2} \Phi_x^2 
         - \frac{\ve}{\mu}\frac{ [ G(\ve\zeta)\Phi + \ve\mu \zeta_x \Phi_x ]^2 }{2 (1+ \ve^2 \mu \zeta_x^2)} = 0.
\end{array}
\right.
\end{align} 
Given a solution of this system, we reconstruct the potential $\phi$ by solving the Laplace
equation in the domain $\Omega_t$ (cf. \cite{Lannes,Wu1}), and then define the average velocity in the context
of the full water-wave problem by
\begin{equation}\label{averaged}
\bar{V}(x,t) = \frac{1}{1+\ve\zeta}\int_{-1}^{\ve\zeta} \partial_x \phi(x,z,t)dz.
\end{equation}
From \cite{LannesBOOK} (Thm 4.16), we have the following result.
\begin{theorem}

\noindent
(1) For large enough $s$, there exists a unique solution $(\zeta,\Phi) \in C(0,T/\ve, H^{s} \times \dot{H}^{s+1})$ 
of the ZCS water-wave system.

\noindent
(2) For the average velocity, we have $\bar{V} \in C(0,T/\ve, H^{s-3})$.
\end{theorem}

In the shallow-water small-amplitude regime specified aboce ($\mu \ll 1$, $\ve = \scrO(\mu)$), 
one can derive the Peregrine system. 
For one dimensional surfaces and flat bottoms, these equations couple the free surface
elevation $\zeta$ to the vertically averaged horizontal component of the velocity,
and can be written as
\begin{equation}\label{Peregrine}
	\left\lbrace
	\begin{array}{l}
	\xi_t+\big[(1+\ve\xi)u\big]_x=0,\\
	u_t+\xi_x+\ve u u_x =\frac{\mu}{3}u_{xxt}.
	\end{array}\right.
\end{equation}
%where $O(\mu^2)$ terms have been discarded.\\
Based on results proved in \cite{Amick,Schonbek}, the authors of \cite{BCSII} 
formulate the following result showing that this system is globally well posed.
\begin{theorem}
Suppose $s \ge 1$, and initial data $(\xi_0,u_0) \in H^s \times H^{s+1}$ are given
with the additional assumption that $\inf{\xi} > -1$.
Then there is a unique solution $(\xi,u)$ of the system \eqref{Peregrine}
which for any $\mathcal{T}>0$ lies in $C(0,\mathcal{T},H^s(\R)) \times C(0,\mathcal{T},H^{s+1}(\R))$, 
and such that $\xi(x,0) = \zeta_0(x)$ and $u(x,0) = u_0(x)$.
Moreover the solution depends continuously on the initial data in the norm
of $C(0,\mathcal{T},H^s(\R)) \times C(0,\mathcal{T},H^{s+1}(\R))$.
\end{theorem}

Proving convergence of the unknown quantities in the model system
to the unknowns in the full water-wave problem requires consistency
and stability. Stability of the Peregrine system was proved
in Proposition 6.5 in \cite{LannesBOOK}. 
Restricting this result to a flat bottom and to one space dimension
yields the following theorem.
\begin{theorem}
If a pair of functions $(\txi,\tu)$ exists, such that
\begin{equation*}%\label{eq1nond}
	\left\lbrace
	\begin{array}{l}
	\txi_t+\big[(1+\ve\txi) \tu \big]_x=  r, \\
	\tu_t - \frac{\mu}{3}u_{xxt} + \txi_x + \ve \tu \tu_{\tx}  =  R,
	\end{array}\right.
\end{equation*}
then 
\begin{equation}\label{Peregrine-convergence}
\big\| \big[ (\txi,\tu) - (\xi,u) \big](\cdot,t) \big\|_{H^{s}\times H^{s}_{\mu}} 
\le C \Big( \big\| \big[ (\txi,\tu) - (\xi,u) \big](\cdot,0) \big\|_{H^{s}\times H^{s}_{\mu}} + t \|(r,R)\|_{L^{\infty}} \Big),
\end{equation}
where the norm $\|\cdot\|_{H^{s}_{\mu}}$ is defined by
$\|f\|_{H^{s}_{\mu}}^2 = \|f\|_{H^{s}}^2 + \mu \|f_x\|_{H^{s}}^2$.
\end{theorem}
\noindent
Next following the procedure laid out in \cite{BCL,ConstantinLannes,LannesBOOK}, we define
consistency based on system \eqref{Peregrine}.
\begin{definition}\label{Def1}
A family of function pairs $(\zeta^{\ve,\mu}, v^{\ve,\mu})$ is consistent with
\eqref{Peregrine} if for all $\ve >0$, we have
\begin{equation*}%\label{eq1nond}
\left\lbrace
\begin{array}{l}
\zeta^{\ve,\mu}_t+\big[(1+\ve\zeta^{\ve,\mu}) v^{\ve,\mu} \big]_x=  \ve^2 r^{\ve,\mu}, \\
v^{\ve,\mu}_t - \frac{\mu}{3}v^{\ve,\mu}_{xxt} + \zeta^{\ve,\mu}_x + \ve v^{\ve,\mu} v^{\ve,\mu}_{x}  =  \ve^2 R^{\ve,\mu},
\end{array}\right.
\end{equation*}
with $(r^{\ve,\mu},R^{\ve,\mu})$ bounded in $L^{\infty}(0,T/\ve, H^s(\R) \times H^s(\R) )$.
\end{definition}
\noindent
Clearly, if we are able to find a family of function pairs consistent with
the Peregrine system, then by the stability result, these functions will
converge towards the unknowns of the Peregrine system.
It turns out that both the ZCS equations and the KdV equation can be shown
to be consistent with the Peregrine system.
From Corollary 5.20 in \cite{LannesBOOK} (with flat bottom),
we have the following result.
\begin{theorem}
\noindent
The water-wave equations are consistent with the Peregrine system. 
Indeed for a solution $\zeta,\Phi$ of the water wave problem, we 
can define $\phi$ and $\vbarWW$ as explained above, and we have
\begin{equation*}%\label{eq1nond}
	\left\lbrace
	\begin{array}{l}
	\zeta_t+\big[(1+\ve\txi) \vbarWW \big]_x=  0, \\
	{\vbarWW}_t - \frac{\mu}{3}{\vbarWW}_{xxt} + \zeta_x + \ve \vbarWW {\vbarWW}_{\tx}  =  \mu^2 R,
	\end{array}\right.
\end{equation*}
%with $R(\cdot,t) \in L^{\infty}\big( 0, T / \ve, H^N(\R) \big)$
with $\|R(\cdot,t)\|_{H^s}$ bounded for $ t \in [0,T/\ve]$.
\end{theorem}
\noindent
Now following the proof of Corollary 6.23 in \cite{LannesBOOK},
one may put all these theorems together to obtain the following result.
\begin{theorem}
Suppose initial data $(\zeta_0,\Phi_0) \in H^N(\R) \times \dot{H}^{N+1}(\R)$ 
are given for a large enough Sobolev index $N$.
Defining initial data for \eqref{Peregrine} by $\xi_0 = \zeta_0$
and $u_0 = \frac{1}{1+\ve \zeta_0}\int_{-1}^{\ve\zeta_0} \partial_x \phi(x,z,0)dz$,
there exists a constant $C$, such that the estimate
\begin{equation}\label{MainEstimate}
\|(\zeta,\bar{V}) - (\xi,u)\|_{L^{\infty}} \le C \mu^2 t
\end{equation}
holds for the solutions of \eqref{Z-C-S} and \eqref{Peregrine}
and with $\vbarWW$ defined by \eqref{averaged}.
\end{theorem}
\noindent
Next we turn to the uni-directional KdV model. Existence, uniqueness and continuous
dependence on the initial data follow from the results proved in \cite{BonaSmith},
and are by now classical.
\begin{theorem}\label{KdV-existence}
For the KdV equation with initial data in $H^s$, where $s>1$, there is a unique
solution $\eta \in C(0,\mathcal{T},H^s)$ for ant $\mathcal{T} > 0$, 
and the solution depends continuously on the initial data.
\end{theorem}
\noindent
Note also that it was proved in \cite{Israwi} that $\eta \in C^1(0,T/\ve,H^{s-1})$.
In order to prove consistency in the sense of Definition \ref{Def1},
we need to define an appropriate velocity. Following \cite{Israwi}, 
we define
\begin{equation}\label{KdV-velocity}
v_{KdV} = \eta - \frac{\ve}{4} \eta^2 - \frac{\mu}{6} \eta_{xt}.
\end{equation}
Now given initial data $(\xi_0,u_0) \in H^s\times H^{s+1}$
and the solution $(\xi,u) \in C(0,\mathcal{T},H^s \times H^{s+1})$ of 
the system \eqref{Peregrine}
we define initial data for \eqref{KdV} by $\eta_0  = \xi_0$,
and using the solution $\eta \in C(0,\mathcal{T},H^s)$ guaranteed by Theorem
\ref{KdV-existence} we define $v_{KdV}$ by \eqref{KdV-velocity}.
Then following the proof laid out in \cite{Israwi}, we can obtain
the following estimate.
\begin{theorem}
With the above previous, we have the estimate
\begin{equation}\label{KdV-convergence}
\|(\eta,v_{KdV}) - (\xi,u)\|_{L^{\infty}} \le C \mu^2 t.
\end{equation}
\end{theorem}
Thus it becomes clear that since the Peregrine system approximates the full water wave problem,
so do solutions of the KdV equation.
\section{Convergence of momentum density}
We now give a proof of Theorem 1.
\begin{proof}
First of all, from \eqref{Peregrine-convergence} and \eqref{KdV-convergence}
and the triangle inequality we have
\begin{equation}\label{KdV-Euler-velocity-surface}
\|\vbarWW - v_{KdV} \|_{L^{\infty}} \le C \mu^2 t \quad \mbox{ and } \quad
\|\zeta   - \eta  \|_{L^{\infty}}  \le C \mu^2 t.
\end{equation}
Next we have
\begin{eqnarray*}
\int_{-1}^{\ve \zeta} \phi_x dz - I 
& = & 
 \int_{-1}^{\ve \zeta} \phi_x dz - \frac{1}{1+\ve\zeta} \int_{-1}^{\ve \zeta} \phi_x dz 
%& &
 +  \frac{1}{1+\ve\zeta} \int_{-1}^{\ve \zeta}\dx \phi dz -
        \Big( \eta - \frac{\ve}{4}\eta^2 - \ve\frac{1}{6} \eta_{xt}\Big) \\
& &   + \Big( \eta - \frac{\ve}{4}\eta^2 - \ve\frac{1}{6} \eta_{xt}\Big)
    -   \Big( \eta + \frac{3\ve}{4}\eta^2 + \ve\frac{1}{6} \eta_{xt}\Big) \\
& = &  \Big(1- \frac{1}{1+\ve \zeta} \Big) \int_{-1}^{\ve \zeta} \phi_x dz + \vbarWW - v_{KdV}
  - \ve \Big(\frac{1}{4} + \frac{3}{4} \Big) \eta^2 
  - \frac{\mu}{6} \big(\eta_{xt} + \eta_{xx} \big). \\
\end{eqnarray*}
Now observe that equation \eqref{KdV} together with the fact that
solutions of \eqref{KdV} are bounded in $H^s(\R)$ for all time
(cf. \cite{BonaSmith})
can be used to show that $ \|\eta_{xx}(\cdot,t) + \eta_{xt}(\cdot,t)\|_{L^{\infty}} \le C \mu$.
Using this estimate together with the triangle inequality, and the fact that
$ \|\vbarWW(\cdot,t) - v_{KdV}(\cdot,t)\|_{L^{\infty}} \le C \mu^2 t$ 
from \eqref{KdV-Euler-velocity-surface} leads to
\begin{equation*}
\Big\| \int_{-1}^{\ve\zeta(\cdot,t)}\dx \phi(\cdot,z,t)dz - I(\cdot,t) \Big\|_{L^{\infty}}  \\
\leq  \ve \| \zeta(\cdot,t) \vbarWW(\cdot,t)  -\eta^2(\cdot,t)  \|_{L^{\infty}} + C \mu^2 (1+t).
%\leq  \ve \Big| \zeta  \int_{-1}^{\ve \zeta} \phi_x dz  -\eta^2  \Big| + C \mu^2 (1+t).
\end{equation*}
Finally, notice that 
\begin{eqnarray*}
\zeta \vbarWW - \eta^2 
& = & \zeta^2 - \zeta^2 + \zeta \vbarWW - \eta^2  \\
& = & (\zeta + \eta)(\zeta - \eta) + \zeta (-\zeta + \vbarWW) \\
& = & (\zeta + \eta)(\zeta - \eta) + \zeta (\vbarWW - v_{KdV} + v_{KdV} -\eta + \eta -\zeta).
\end{eqnarray*}
The function $\zeta$ is bounded in $C(0,T/\ve,H^s)$, 
$\vbarWW$ is bounded in $C(0,T/\ve,H^{s-3})$
$\eta$ is bounded in $C(0,\mathcal{T},H^s)$
and $v_{KdV}$ is bounded in $C(0,T/\ve,H^{s-2})$,
so that we have
\begin{equation*}
\Big\| \int_{-1}^{\ve\zeta(\cdot,t)}\dx \phi(\cdot,z,t)dz - I(\cdot,t) \Big\|_{L^{\infty}}  \\
\leq  \ve C \Big\{
  \| \zeta - \eta \|_{L^{\infty}}
+ \| \vbarWW - v_{KdV} \|_{L^{\infty}}
+ \|v_{KdV} - \eta \|_{L^{\infty}}
+ \| \eta - \zeta \|_{L^{\infty}}
%|\zeta(\cdot,t) - \eta(\cdot,t) \|_{L^{\infty}}
%+ |\vbarWW(\cdot,t) - v_{KdV}(\cdot,t) \|_{L^{\infty}}
%+ |v_{KdV}(\cdot,t) - \eta(\cdot,t) \|_{L^{\infty}}
%+ |\eta(\cdot,t) - \zeta(\cdot,t) \|_{L^{\infty}}
\Big\}.
\end{equation*}
Finally using \eqref{KdV-Euler-velocity-surface}, and the definition of $v_{KdV}$
\eqref{KdV-velocity} along with the fact 
that $\eta \in C^1(0,T/\ve,H^{s-1})$ (cf. \cite{Israwi}) proves the required result.
\end{proof}

\vskip 0.05in
\noindent
{\bf Acknowledgments.}
This research was supported by the Research Council of Norway under grant no. 239033/F20.

%}

\end{document}